\documentclass[sigconf, screen, nonacm]{acmart}
\AtBeginDocument{%
  }

\copyrightyear{2025}
\acmYear{2025}
\setcopyright{cc}
\setcctype{by-nc}
\acmConference[CUI '25]{Proceedings of the 7th ACM Conference on Conversational User Interfaces}{July 8--10, 2025}{Waterloo, ON, Canada}
\acmBooktitle{Proceedings of the 7th ACM Conference on Conversational User Interfaces (CUI '25), July 8--10, 2025, Waterloo, ON, Canada}
\acmDOI{10.1145/3719160.3736606}
\acmISBN{979-8-4007-1527-3/2025/07}

\usepackage{float}
\usepackage{subcaption}

\usepackage{multirow}

\definecolor{LLM-generated}{RGB}{72, 120, 208}
\definecolor{LLM-pretested}{RGB}{238, 133, 74}

\begin{document}

\title{Exploring LLMs for Automated Generation and Adaptation of Questionnaires}
\thanks{Preprint. This paper has been accepted to ACM Conversational User Interfaces
2025 (CUI '25); please cite \url{https://doi.org/10.1145/3719160.3736606} instead.}

\author{Divya Mani Adhikari}
\email{diad00001@stud.uni-saarland.de}
\orcid{0009-0001-6820-489X}
\affiliation{%
  \institution{Interdisciplinary Institute for Societal Computing}
  \institution{Saarland University}
  \city{Saarbrücken}
  \state{}
  \country{Germany}
}

\author{Alexander Hartland}
\email{alexander.hartland@uni-saarland.de}
\orcid{0000-0003-4842-3881}
\affiliation{%
  \institution{Department of European Social Research}
  \institution{Saarland University}
  \city{Saarbrücken}
  \state{}
  \country{Germany}
}
\additionalaffiliation{%
  \institution{Interdisciplinary Institute for Societal Computing, }
  \institution{Saarland University}
  \city{Saarbrücken}
  \state{}
  \country{Germany}
}

\author{Ingmar Weber}
\email{iweber@cs.uni-saarland.de}
\orcid{0000-0003-4169-2579}
\affiliation{%
  \institution{Interdisciplinary Institute for Societal Computing}
  \institution{Saarland University}
  \city{Saarbrücken}
  \state{}
  \country{Germany}
}

\author{Vikram Kamath Cannanure}
\orcid{0000-0002-0944-7074}
\email{cannanure@cs.uni-saarland.de}
\affiliation{%
  \institution{Interdisciplinary Institute for Societal Computing}
  \institution{Saarland University}
  \city{Saarbrücken}
  \state{}
  \country{Germany}
}

\renewcommand{\shortauthors}{Adhikari et al.}

\begin{abstract}
Effective questionnaire design improves the validity of the results, but creating and adapting questionnaires across contexts is challenging due to resource constraints and limited expert access. Recently, the emergence of LLMs has led researchers to explore their potential in survey research. In this work, we focus on the suitability of LLMs in assisting the generation and adaptation of questionnaires. We introduce a novel pipeline that leverages LLMs to create new questionnaires, pretest with a target audience to determine potential issues and adapt existing standardized questionnaires for different contexts. We evaluated our pipeline for creation and adaptation through two studies on Prolific, involving 238 participants from the US and 118 participants from South Africa. Our findings show that participants found LLM-generated text clearer, LLM-pretested text more specific, and LLM-adapted questions slightly clearer and less biased than traditional ones. Our work opens new opportunities for LLM-driven questionnaire support in survey research.

\end{abstract}


\begin{teaserfigure}
\centering
  \includegraphics[width=0.7\textwidth]{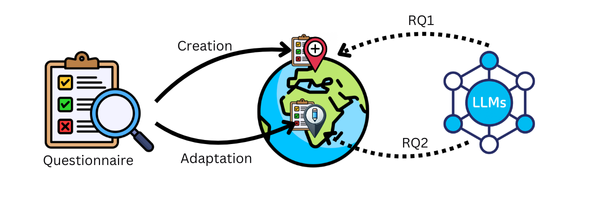}
  \caption{We explore the role of LLMs in questionnaire design by understanding their capabilities in creating and pretesting new questionnaires (RQ1: creation) and modifying existing ones (RQ2: adaptation) to suit different contexts.}
  \Description{This figure illustrates the high-level research workflow investigating the interplay between large language models (LLMs) and questionnaire-based studies across diverse geographical contexts. The process begins with the creation and adaptation of a questionnaire, tailored for various regional and cultural settings (depicted by the globe and location pins). Two primary research questions (RQ1 and RQ2) guide the inquiry: RQ1 examines the potential of LLMs in the creation of contextually relevant questionnaires, while RQ2 investigates how well LLMs can adapt existing questionnaires to new cultural and linguistic settings. The dotted arrows represent the analytical feedback loop between the human-designed questionnaires and the computational capabilities of LLMs, aiming to enhance the global applicability and robustness of survey-based research methods.}
  \label{fig:teaser}
\end{teaserfigure}


\maketitle

\section{Introduction}
Designing well-structured questionnaires is essential for gathering survey data, as it influences the validity of collected responses~\cite{lenzner_effects_2012, knauper_question_1997} from diverse populations worldwide. While standardized surveys exist, designing a new questionnaire remains challenging, requiring a sequence of logically structured questions aligned with specific research objectives~\cite{taherdoost_designing_2022, synodinos_art_2003, jenn_designing_2006, boynton_selecting_2004}. Researchers must craft clear, relevant, and precise questions to extract meaningful insights without causing confusion or misinterpretation among respondents~\cite{lenzner_effects_2012}. Pretesting is a crucial step in questionnaire design, ensuring that errors, biases, or ambiguities—especially in sensitive~\cite{https://doi.org/10.1046/j.1365-2648.2003.02579.x, Lenzer2024-cn} or cross-cultural~\cite{Gjersing2010, 00007632-200012150-00014} surveys—are identified before deployment. Pretesting is necessary for cross-cultural studies~\cite{behr_web_2017, Blair2005}, global market research~\cite{reynolds_pre-testing_1993}, and international policy assessments~\cite{doi:10.1177/0269216318818299}, where question phrasing and interpretation can vary significantly across different regions. However, resource constraints~\cite{stahura_methods_2005}, participant availability~\cite{johanson_initial_2010,blair_sample_2011}, and limited expert access~\citep{doi:10.1177/1525822X10379795} make pretesting difficult for new questionnaires or adapting existing questionnaires to different contexts. 

Automated pre-testing with Large Language Models (LLMs) offers a promising solution to key challenges in survey design~\cite{jansen_employing_2023}. LLMs have been widely adopted for text generation and analysis due to their ability to produce complex, coherent responses based on given instructions~\cite{liang2024controllabletextgenerationlarge}. Additionally, LLMs can simulate pilot studies by adopting different personas~\cite{sarstedt_using_2024}, allowing researchers to evaluate survey effectiveness before deployment. Furthermore, LLMs can analyze questionnaires to detect potential issues, enhancing the pre-testing process and reducing the need for multiple human iterations~\cite{olivos_chatgptest_2024, sarstedt_using_2024}. This work integrates LLMs (specifically GPT-4o) into a questionnaire design to understand their utility and perform a preliminary evaluation. Through our work, we answer the following research questions: 

\begin{description}
    \item[RQ1: Creation] How suitable are LLMs for automating new questionnaire generation and pretesting?
    \item[RQ2: Adaptation] How suitable are LLMs at adapting existing questionnaires for a specific target audience?  
\end{description}

\begin{table}[!htb]
    \centering
    \begin{tabular}{|p{0.15\linewidth}|p{0.18\linewidth}|p{0.12\linewidth}|p{0.4\linewidth}|}
    \hline
       RQ  & No. of participants & Country & Subject expertise requested  \\
       \hline
       \multirow{2}{*}{Creation}  & \shortstack[l]{\\238 \\ (Prolific)} & United States & Politics, Psychology, Social Work, Sociology \\
       \cline{2-4}
        & \shortstack[l]{\\13 \\ (Experts)}  & EU & Political Science, Psychology, Social Science \\
        \hline
       Adaptation & \shortstack[l]{\\118 \\ (Prolific)} & South Africa & Earth, Environment, or Climate Sciences, Politics, Psychology, Social Work, Sociology \\
       \hline
    \end{tabular}
    \caption{The number of participants, their geographic distribution, and subject backgrounds for two research questions: Creation and Adaptation. Participants were recruited from Prolific and expert networks across different regions (US, EU, and South Africa). For each study, participants aged 18+ with an undergraduate degree in the listed subjects were recruited.}
\label{tab:participants-summary}
\end{table}

To address these research questions, we conducted two studies using Prolific\footnote{\url{https://prolific.com/}}, an online crowdsourcing platform. For RQ1, we utilized our pipeline to generate a questionnaire on political trust and pretested it with 238 Prolific participants and 13 experts. For RQ2, we adapted a U.S.-centric standardized public opinion survey on climate change for a South African audience and gathered feedback from 118 South African participants on Prolific. As part of our evaluation, participants rated LLM-generated and LLM-pretested survey questions on a 5-point Likert scale based on clarity, bias, relevance, and specificity. They then compared both types of questions side by side, selecting the one they found clearer and briefly explaining their choice.

For our first research question (RQ1: Creation), we found that LLMs have the potential to generate relevant and specific questions while lacking in pretesting survey questionnaires. This suggested that LLMs are already good at generating high-quality questions, and further pretesting might be unnecessary. For the second question (RQ2: Adaptation), participants on Prolific found the pretested questionnaire slightly more explicit, unbiased, and relevant, and thus, more appropriate for a South African target audience. This result suggested that LLMs can adapt existing questionnaires to a specific target audience.

Overall, we make the following contributions:
\begin{itemize}
   
    \item We show empirical results from two studies on Prolific about the suitability and challenges of LLMs in creating and adapting survey questionnaires.
    
    \item We introduce a novel LLM-based pipeline designed for the automated generation and pretesting of survey questionnaires using an LLM-simulated Pilot study. The pipeline can be adapted for use with any existing LLM or in any domain.
    
    \item We discuss strategies for questionnaire design for clarity, specificity, and neutrality in survey questions.
\end{itemize}

\section{Related Work}
\subsection{Automated Survey Administration}
Existing research has primarily focused on automating survey administration using computational techniques, particularly conversational systems. Chatbots have shown promise in replacing human interviewers while improving data quality~\cite{kim_comparing_2019}, increasing completion rates and reducing costs~\cite{ndashimye_effectiveness_2024, fei_automated_2022}, and even enhancing interaction~\cite{abbas_university_2021}. More recently, Large Language Models (LLMs) have further expanded the automation possibilities in survey research. LLM-powered interviewers can collect data at a quality comparable to traditional methods while offering scalability ~\cite{wuttke2024aiconversationalinterviewingtransforming}. Moreover, ~\citet{villalba_automated_2023} showed LLM-augmented chatbots improve user experience through dynamic follow-up question generation during interviews. While automation in survey administration is well-studied, the automation of questionnaire generation and pretesting remains underexplored. With the rise of LLMs, there is growing interest in leveraging their text generation and analytical capabilities to enhance survey research. \citet{jansen_employing_2023} discuss how LLMs could support the pretesting of survey instruments, signaling a shift toward broader automation in survey design.

\subsection{Questionnaire Generation}
Questionnaires consist of a series of questions that include coreferences and inherent constraints, making their creation more complex than generating general sequential questions~\cite{lei_qsnail_2024}. Existing research on question generation has primarily focused on creating independent, factual questions ~\citep{rajpurkar_know_2018, xiao_ernie-gen_2020}. This approach has been adapted within survey research to generate follow-up questions in chatbot-administered interviews~\cite{su_follow-up_2018, rao_s_b_improving_2021, ge_what_2023, seltzer_smartprobe_2023}. Beyond individual question generation, researchers have also explored methods for generating sequences of questions~\cite{gao_interconnected_2019, chai_learning_2020}. However, work specifically on questionnaire generation remains limited. Early research by \citet{jenkins_automating_2000} explored automating questionnaire design by reusing standardized questions and searchable question libraries. More recently, studies have investigated the potential of LLMs in this space\cite{maiorino_application_2023, parker_flexibility_2023}. LLMs have been used in generating rephrases of questionnaires~\cite{zou_pilot_2024}, generating diverse versions of existing questionnaires to reduce respondent fatigue in longitudinal studies~\cite{yun_keeping_2023}, and even generating HR-related questionnaires~\cite{laraspata_enhancing_nodate}. Similarly, \citet{rothschild_opportunities_2024} found using LLMs in the survey designing process substantially reduces the survey launch time. However, \citet{lei_qsnail_2024} and \citet{laraspata_enhancing_nodate} found that LLM-generated questions were generally too broad, had generic wording, and lacked specificity. \citet{laraspata_enhancing_nodate} propose the use of Retrieval-Augmented Generation (RAG)~\cite{NEURIPS2020_6b493230} as a strategy to reduce hallucinations and enhance the specificity of generated questions.

\subsection{Automated Pretesting of Questionnnaire}
Pretesting a questionnaire requires feedback from target participants or domain experts, thus automation in this area remains largely unexplored. Traditional pretesting methods include Cognitive Interviewing~\cite{scott_yes_2020}, Pilot Studies (involving participants)~\cite{ismail_pilot_2018}, Expert Review~\cite{olson_examination_2010} and Behavior Coding (involving domain experts) ~\citep{blair_using_2007, reynolds_pre-testing_1993}. While some early studies explored online adaptations, such as conducting cognitive interviews via Skype or Second Life ~\citep{dean_virtual_2013}, automation in pretesting has remained limited. One exception is Web Probing, a technique used in online surveys where participants are asked follow-up questions (e.g., You answered "Yes"—how did you arrive at this answer?) to understand their thought process ~\citep{lenzner_pretesting_2017, behr_showcasing_2024, neuert_design_2023}.

Recent work has explored using LLMs to automate questionnaire pretesting. \citet{parker_flexibility_2023} used ChatGPT to review interview protocols but found the feedback not critical enough to refine the protocols. Likewise, \citet{olivos_chatgptest_2024, sarstedt_using_2024} demonstrated how ChatGPT can review survey questions to identify potential issues. However, these reviews are limited by the biases of the LLMs and the datasets they have been trained on~\cite{10.1371/journal.pone.0306621, tjuatja_llms_2024}. LLMs can adopt various personas, allowing for the evaluation of questionnaires from diverse perspectives, thereby mitigating biases~\cite{sun_building_2024, guyre_prompt_2024, hohn_using_2024}. Researchers have explored using LLMs to simulate diverse audience perspectives~\cite{jansen_employing_2023,kim_ai-augmented_2023}. \citet{kim2024llmmirrorgeneratedpersonaapproachsurvey} demonstrated that LLMs when initialized with a persona, can simulate human responses on an individual level. Similarly, \citet{argyle_out_2023} showed that properly conditioning GPT-3 enables it to replicate responses from demographic subgroups. 

However, LLMs are not a perfect substitute for human participants. Studies show that while LLMs can infer human biases ~\citep{schramowski_large_2022}, serve as simulated economic agents ~\citep{horton_large_2023}, and generate plausible HCI research data ~\citep{hamalainen_evaluating_2023}, they can also misrepresent and flatten demographic groups ~\citep{wang_large_2024}. Researchers caution against using LLMs as a complete replacement for human responses ~\citep{dominguez-olmedo_questioning_2024, dillion_can_2023}, but suggest they can be valuable supplements in pretesting, particularly in pilot studies and cognitive interviews. \citet{wang_large_2024} also proposed techniques to mitigate misrepresentation issues, supporting the idea that LLMs, while imperfect, can be used to simulate pilot studies and pretesting methods to refine survey design before real-world deployment. Our work considers these techniques and extends existing work by utilizing LLM personas to simulate a Pilot study and analyze questionnaires from a variety of perspectives.

\subsection{Cross-cultural Questionnaire Adaptation}
Adaptation is essential when a questionnaire designed for a specific context, such as a particular language or country, is to be used in a different setting. While no consensus exists for a particular approach~\cite{epstein_review_2015}, much of the existing research has focused on adapting questionnaires from one language to another through translation~\cite{beaton_guidelines_2000,haavisto_questionnaires_2024}. The adapted questionnaire requires further validation, such as pretesting, before being deployed to a new audience~\cite{epstein_review_2015,beaton_guidelines_2000}. \citet{haavisto_questionnaires_2024} translated a questionnaire and used GPT-4 to evaluate the translation quality to suggest improvements. They found that GPT-suggested changes can achieve similar results as conventional methods.

Even within the same language, such as English, adaptation is crucial to mitigate potential misunderstandings due to cross-cultural differences. \citet{harzing_does_2005} found minimal difference in the participant responses when adapting an English questionnaire; however, \citet{sousa_questionnaire_2017} suggested that participants can interpret original and adapted questions differently, with adaptation generally making the questions more straightforward to understand. While existing works rely on expert feedback or back-translation for questionnaire adaptation~\cite{EPSTEIN2015360}, our work explores questionnaire adaptation and evaluation using simulation of participants from the new target audience through a Pilot study using LLMs.
\smallbreak
\textit{In summary}, existing work has automated survey administration using chatbots and LLMs, improving scalability and data quality, but questionnaire generation and pretesting remain underexplored. While LLMs aid in structuring and adapting surveys, issues of specificity and bias persist. Our research aims to bridge these gaps by enhancing questionnaire generation through RAG and simulating pilot studies for automated pretesting and adaptation, to ultimately improve the quality and relevance of survey instruments.

\section{Methodology}

\begin{figure*}[ht!]
    \centering
    \includegraphics[width=\linewidth]{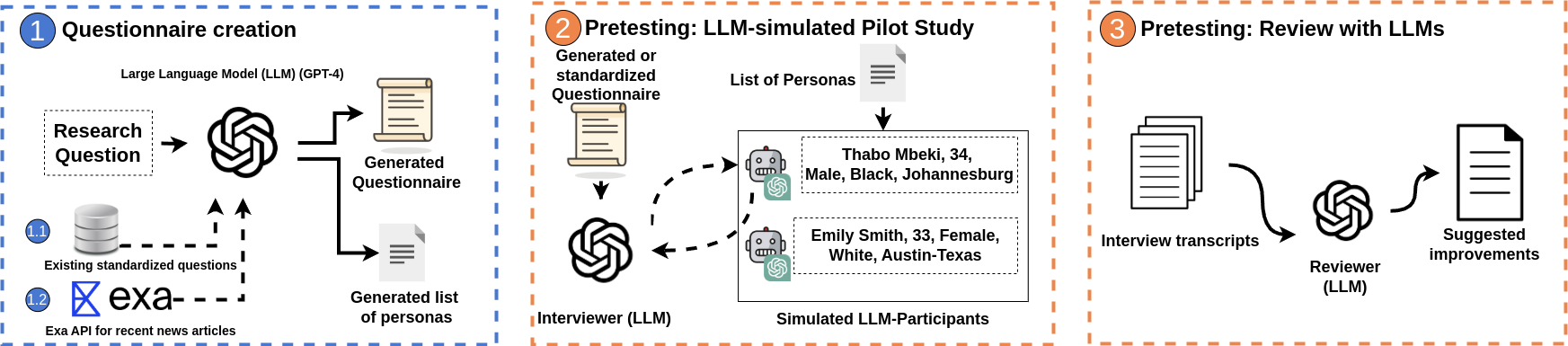}
    \caption{Our complete questionnaire creation and pretesting pipeline involve: (1) Using the research question and relevant contextual information, such as a few questions from existing standardized questionnaires and a summary of recent relevant news articles (see figure \ref{fig:rag_internals}), the LLM generates a questionnaire and a list of Personas. (2) The generated question is then pretested through a simulated pilot study using the generated personas. (3) Interview transcripts from the simulated pilot study are collected and analyzed by the reviewer LLM, which provides suggestions for improving the questionnaire.}
    \label{fig:qgen_pretest_review}
    \Description{The figure shows the 3 stages in our questionnaire generation and pretesting pipeline. The first stage shows the questionnaire generation stage. The research question along with relevant questions from standardized surveys as well as a summary of recent relevant news articles is sent to the LLM which generates a questionnaire and a list of Personas. The second stage shows the questionnaire pretesting stage where an LLM acts as an interviewer and multiple LLMs act as participants in a simulated Pilot study. The interviewer LLM conducts the interview based on the previously generated questionnaire and the transcripts of the interviews are collected. In the final stage, the figure shows a Reviewer LLM taking the collected interview transcripts and returning suggestions to improve the questionnaire.}
\end{figure*}
\subsection{System Design}

\subsubsection{Questionnaire Generation}
\label{ref:questionnaire-generation}
\begin{figure}[ht]
    \centering
    \includegraphics[width=0.7\linewidth]{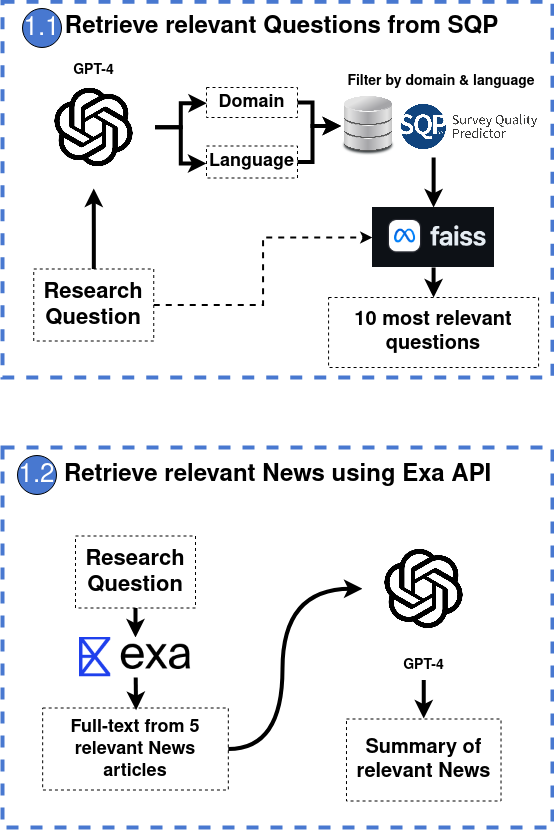}
    \caption{(1) Based on the research question, the domain and the language of the questionnaire are detected which is used to filter questions from the SQP database. Using FAISS, the 10 most relevant questions are retrieved using Cosine similarity. (2) We used the Exa API which returns the full text of 5 relevant News articles based on the research question which is then summarized by GPT-4. Both the relevant questions and summary of relevant News are included in the Questionnaire generation prompt.}
    \Description{The figure shows two blocks which are the subcomponents of the questionnaire generation stage used for the Retrieval Augmented Generation. The first block shows how GPT-4 determines the domain and language of the research question which is then used to filter questions from the Survey Quality Predictor database. The filtered questions are then ranked using Cosine similarity to retrieve the top 10 relevant questions. The second block shows how relevant news articles are retrieved based on the given research question using the Exa platform. The retrieved news articles are summarized using GPT-4 and included in the questionnaire generation prompt.}
    \label{fig:rag_internals}
\end{figure}

We start with a detailed research question the researcher wants to answer through the survey. The research question might include the goal of the study, details about the target audience, and what specific topics they want the questionnaire to include. The researcher submits the research question along with the number of questions that need to be generated. In addition to the research question, the LLM receives relevant contextual information supporting questionnaire generation: a summary of recent news articles and relevant standardized questions through the questionnaire generation prompt. We use the Exa API \footnote{\url{https://exa.ai}} to retrieve 5 recent relevant News articles and then use GPT-4 to summarize them. We also use 10 relevant questions retrieved from the Survey Quality Predictor (SQP) database\footnote{\url{https://sqp.gesis.org/}} \cite{sqp}. SQP is a multilingual and multi-domain database of curated and crowdsourced questions from existing surveys. First, we determine the domain and language of the questionnaire from the given research question and then use this information in filtering a subset of SQP questions. We then use FAISS~\cite{douze2024faiss} to rank these questions and retrieve the 10 most relevant questions using Cosine similarity as shown in figure \ref{fig:rag_internals}. This contextual information could make the generated questionnaires more relevant to current affairs and the political climate and specific to the research objective.

Along with the questionnaire, the LLM also generates a list of personas, which are descriptions of fictional people of the target audience. Each persona description includes essential characteristics about the person, such as their name, gender, age, employment, etc., and dynamic preferences based on the research question. For example, if the research question is to study the perspectives of people in Germany about the use of AI in the workplace, the generated list of personas would include fictional descriptions of people from Germany where some might encourage the usage of AI in the workplace while others might oppose it or have mixed opinions about it. This list of personas is then used during the pretesting phase to simulate the participants for the Pilot study.

\subsubsection{Questionnaire Pretesting}
The generated questionnaire is then pretested by simulating a real-world pilot study using multiple LLM instances. One of the LLMs is initialized as the interviewer, and multiple LLMs are initialized as the participants based on the previously generated list of personas. The interviewer LLM then asks questions from the generated questionnaire, and the simulated participant LLMs respond accordingly to their persona descriptions. After each question, the interviewer LLM asks follow-up questions, dynamically generated based on the previous responses of the participant. The generated follow-up questions followed the cognitive interviewing technique to determine the participants' thought processes. The conversation history (interview transcript) between the interviewer LLM and the participant LLMs is stored for further analysis.

After the pilot, the collected interview transcripts are analyzed by a reviewer LLM, which is initialized to find potential issues in the pretested questionnaire, such as double-barreled, leading, or biased questions. The reviewer LLM gets the initial research questions, the questionnaire under pretest, and the interview transcript and persona description of each simulated interview as context data. The reviewer LLM then generated a review with four topics: the question with a problem, the problem it found, the participants who faced the situation, and the suggested changes to the question (see table \ref{table:reviewer_example}). The complete pipeline is shown in Figure \ref{fig:qgen_pretest_review}. The prompts used in the pipeline are shown in Appendix \ref{appendix:prompts}.
\\

\noindent\fbox{\begin{minipage}{0.45\textwidth}
{\fontfamily{cmtt}\selectfont\small
Statement: P1. The US Congress acts in the best interests of the citizens of the United States.\\
Problem: Participants like Persona 16 had difficulty with this statement, expressing uncertainty due to its generality and potential overlap with political interests, suggesting ambiguity.\\
Sources: Interview 16, Participant Sofia Garcia.\\
Suggestion: The US Congress consistently prioritizes citizens' best interests over political agendas.}
\end{minipage}
}
\captionof{table}{Example feedback from the reviewer in the pretesting of the Political Trust questionnaire.}
\label{table:reviewer_example}

\subsection{Formative Fesability Studies}

To assess the feasibility of our pipeline, we conducted two formative studies: one for questionnaire generation and another for pretesting. In the \textit{questionnaire generation} feasibility study, we defined several key parameters: the generation prompt, the inclusion of existing questions from the SQP database, summaries from recent relevant news articles, temperature, and random seed. For each combination of these parameters, we generated a questionnaire and evaluated its \textit{relevance} and \textit{specificity} using the metrics defined by \citet{lei_qsnail_2024}. Our findings revealed that adding more context increased question specificity but often reduced relevance to the original research objective. For \textit{pretesting}, we replicated the 2016 GESIS pretest study on the ICT usage at work module~\cite{Meitinger2016-hs} using our LLM-based pilot study simulation.  Our pipeline detected issues in 5 out of the 13 tested questions. These results reinforced the feasibility of our questionnaire generation and pretesting pipeline, allowing us to conduct future experiments.

\subsection{Evaluation Metrics}

\begin{table}[htb!]
\small
\centering
\begin{tabular}{|p{0.75\linewidth}|p{0.15\linewidth}|}
\hline
\textbf{Statement} & \textbf{Criterion} \\ \hline
The $\langle \text{statement/question} \rangle$ is ambiguous.        & \multirow{2}{*}{Clarity} \\ \cline{1-1}
The $\langle \text{statement/question} \rangle$ is clear.        &                            \\ \hline
The $\langle \text{statement/question} \rangle$ is biased/leading.        & \multirow{2}{*}{Bias} \\ \cline{1-1}
The $\langle \text{statement/question} \rangle$ is impartial.        &                            \\ \hline
The $\langle \text{statement/question} \rangle$ is relevant to $\langle \text{topic} \rangle$.        & \multirow{2}{*}{Relevance} \\ \cline{1-1}
The $\langle \text{statement/question} \rangle$ has no connection to $\langle \text{topic} \rangle$.        &                            \\ \hline
The $\langle \text{statement/question} \rangle$ asks something specific about $\langle \text{topic} \rangle$.        & \multirow{2}{*}{Specificity} \\ \cline{1-1}
The $\langle \text{statement/question} \rangle$ explores broader aspects beyond just $\langle \text{topic} \rangle$.        &                            \\ \hline
\end{tabular}
\caption{The evaluation statements used in the Prolific studies. Each criterion is assessed directly and also indirectly for validation. We used `question' in $\langle \text{statement/question} \rangle$ in the Cross-cultural study and used `statement' in the Political trust study. Similarly, in $\langle \text{topic} \rangle$, we replaced it with the phrase `the research question.' in the Cross-cultural study, whereas, in the Political trust study, it was replaced with `Political trust.'}
\label{tab:evaluation_statements}
\end{table}

\subsubsection{Evaluation by Prolific users}
To evaluate the LLM-generated and LLM-pretested questions, we asked participants on the Crowdsourcing platform Prolific to rate the initially generated and pretested questions on a Likert scale based on four criteria. We use two criteria defined by \citet{lei_qsnail_2024}: relevance and specificity that measure how relevant is the generated question to a given research question or a topic and how specific the question is within the given general research question, respectively, and use two other common criteria in survey designing: clarity and bias which measure how unambiguous the generated questions are and if the generated questions are leading respectively. For each metric, we used two evaluation statements: one that directly assessed the metric (e.g., "The question is clear.") and another that provided an opposite assessment (e.g., "The question is ambiguous"), both rated by participants on a Likert scale. The evaluation statements for the four criteria are shown in Table \ref{tab:evaluation_statements}.

\subsubsection{Evaluation by Experts}
We also asked domain experts (e.g., faculty members in Political Science and Psychology) who regularly work with survey design to complete the same study as the Prolific participants, i.e., one task to rate generated or pretested questions on the four criteria and the other where they were asked to compare and select the clearer question and provide their reasoning. Then, we analyzed and compared the experts' feedback with the crowdsourcing participants' responses. Expert feedback would provide us with better insights into the effectiveness of LLM-based pretesting.

\begin{figure*}[htb]
    \centering
    \includegraphics[width=0.6\linewidth]{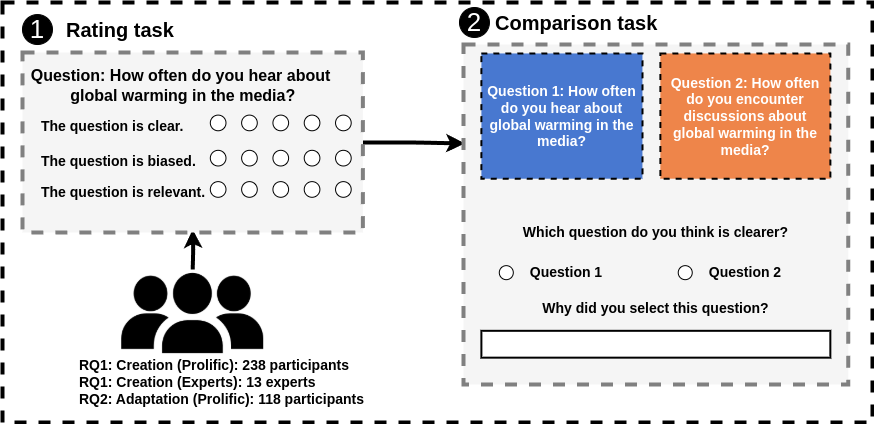}
    \caption{(Simplified) Summary of the tasks in the evaluation studies. (1) In the question rating task, the participants were shown a (randomly selected LLM-generated or LLM-pretested) question and asked to evaluate the question on the four criteria using a 1-5 Likert scale using the 8 evaluation statements shown in Table \ref{tab:evaluation_statements}. (2) In the question comparison task, the participants were (randomly) shown the LLM-generated and the LLM-pretested questions side-by-side and asked to select the question they preferred. They were further asked to explain their decision in a sentence.}
    \Description{The figure shows the simplified version of the two tasks that the crowdsourced workers used to evaluate the questions from the questionnaires. The first task shows a question that is to be rated on our four criteria using a 5-point Likert scale. The second is the comparison task where participants compared LLM-generated and LLM-pretested questions side-by-side, selected the one they preferred, and provided their reason for selecting that question.}
    \label{fig:prolific-evaluation}
\end{figure*}

\subsection{Prolific study design}
The studies on Prolific were designed on Qualtrics\footnote{\url{https://qualtrics.com/}}. At the start, the participants were provided with a detailed description of the purpose of the study. The study procedures were approved by Western German University's ethical review board (ID:[24-09-6]). Then, the participants were given descriptions of the four evaluation criteria: for each criterion, they were given a detailed description, an example of how they could rate the questions, and a test question to check their understanding of the criterion. After this, the participants were led to the main tasks of the study. The participants completed two tasks (rating and comparison) in the study. The first task asked participants to rate questions on the four evaluation criteria. The question was randomly selected from the initially LLM-generated or LLM-pretested list and evaluated by the participants on a Likert scale using eight evaluation statements as shown in Table \ref{tab:evaluation_statements}. Two evaluation statements corresponded to each criterion to mitigate bias and increase the reliability of the measurement. In the second task, the participants were shown a pair of questions (the initially LLM-generated and the LLM-pretested questions) to compare. They were asked to select the one they thought was more explicit for the potential target audience. After this task, the participants were asked to explain their reasoning for choosing that question. The questions and the evaluation statements were randomized in both tasks to mitigate any order effects. The experts in the political trust study were also asked to complete the same survey. The tasks are summarized in Figure \ref{fig:prolific-evaluation}. Example questions from the political trust study are shown in Appendix \ref{appendix:prolific-study-design}.



\subsection{Analysis}

\subsubsection{Quantitative}
For the question rating task, we converted the ratings from the eight evaluation statements into our four evaluation criteria ratings using reverse coding. Then, the mean rating for each criterion was computed. We then computed the difference in the mean ratings for the initially LLM-generated and LLM-pretested questions. To check for statistical significance, we computed the paired samples t-test. The complete summary of the quantitative results is shown in the Appendix in Table \ref{tab:summary}.


\subsubsection{Qualitative}
For the question selection task, we computed the frequency of each question. The frequency indicated the participants' preference regarding which question they perceived as clearer. For the free text responses, we used GPT-4o to do a thematic analysis, categorizing the responses into five themes (shown in Table \ref{tab:thematic_analysis} in the Appendix). Additionally, we manually coded the expert responses to identify the reasons for selecting or rejecting a question (shown in Appendix \ref{appendix:political-trust-study}).

\section{RQ1: Creation: Automating Questionnaire Generation and Pretesting }
\label{section:political-trust}

In this experiment, we evaluated the complete questionnaire generation and pretesting pipeline using LLMs. We chose ``political trust'' because of the long-standing debate about the definition of political trust and its measurement.

\subsection{RQ1: LLM generated questionnaire on Political trust.}
\subsubsection{RQ1: Setup}
We drew inspiration from the political trust module of the World Values Survey (WVS) Wave 7, 2017-2020\cite{Haerpfer2020-uy} to generate the questionnaire on Political trust. The WVS survey features statements under specific questions, rated on a Likert scale. We adopted a similar structure, instructing the LLM to generate statements based on the questions from the WVS political trust module.  The WVS survey, being global, includes placeholders in the questions (e.g., \textit{How do you feel about the $<$name of the national parliament or national congress$>$ in your country?}). We specified the United States as the target country and asked the LLM to generate statements accordingly. The prompt for generating the statements is shown in Appendix \ref{appendix:prompts}. The resulting questionnaire was then pretested with 50 personas generated for the US. The reviewer LLM identified issues with 10 statements in the questionnaire and proposed changes, as detailed in the Appendix in Table \ref{tab:wvs-llm-pretest}. 

\begin{table*}[htb!]
\begin{tabular}{|p{4.5cm}|p{5cm}|p{7.5cm}|}
       \hline
       \textbf{\textcolor{LLM-generated}{Statement generated by LLM}}  & \textbf{\textcolor{LLM-pretested}{Statement after LLM-pretesting}} & \textbf{LLM reasoning} \\
       \hline
        The US Congress acts in the best interests of the citizens of the United States. & The US Congress consistently prioritizes citizens' best interests over political agendas. & Participants like Persona 16 had difficulty with this statement, expressing uncertainty due to its generality and potential overlap with political interests, suggesting ambiguity.\\
        \hline
        The US Congress is effective in passing legislation that benefits the country. & The US Congress efficiently passes legislation with measurable benefits to the country. & Participants such as Persona 15 encountered difficulty deciding on this statement, highlighting the need for clarity or examples of such legislation to guide them, suggesting potential ambiguity.\\
        \hline
        The US Congress consistently demonstrates integrity in its decisions and actions. & The US Congress consistently upholds ethical standards and transparency in its decision-making process. & The statement is vague and seems too conceptual without concrete examples or context. Integrity can be interpreted differently by respondents, and without a clear metric, it's subjective.\\
        \hline
        I feel confident that the US Congress will address the pressing issues of today. & I hold a cautious optimism that the US Congress will adequately tackle today's urgent challenges. & Participant 19 (Ethan Clarke) notes uncertainty and lacks confidence due to inadequate information, indicating the statement might benefit from more context or prompting to be more nuanced.\\
        \hline
        The government's policies are effectively addressing the nation's key problems. & Some government policies effectively address economic, healthcare, and educational challenges. & The phrase "the nation's key problems" is ambiguous as it can be interpreted differently by each respondent, leading to confusion about which problems are being referred to. The response scale does not account for the complexity of addressing multiple issues simultaneously and may not accurately reflect the respondent's views on each issue.\\
        \hline
\end{tabular}
\caption{The statements generated by the Questionnaire generation pipeline (the left column), the same statements after the changes suggested by the LLM-pretesting pipeline (the middle column), and the reasoning from the LLM for that change (the right column) for the Political trust study (RQ1). The remaining five statements are shown in the Appendix in Table \ref{tab:wvs-llm-pretest-remaining}.}
\label{tab:wvs-llm-pretest}
\end{table*}



\subsubsection{RQ1: Participants}
We used Prolific's screener to recruit participants aged 18 and above from the United States with at least an undergraduate degree in Politics, Psychology, Social Work, or Sociology. A total of 238 participants participated in the study, 61\% of which identified as Female. Approximately 41\% of the participants were aged 25 to 34, and about 51\% of the participants leaned left on the Political spectrum. Additionally, 94\% of the participants held at least a University bachelor's degree and reported at least some experience with surveys. About 17\% of the participants mentioned being experienced with the World Values Survey (WVS), which served as the basis for our generated questionnaire. 

For expert feedback, we contacted professors, researchers, and administrative representatives from various research institutes and requested them to share the study link with their teams. We received responses from 13 experts for our study, 46\% of whom identified as Female. About 30\% of the experts were aged 25 to 34, and 85\% held Graduate degrees. Furthermore, 61\% of the experts leaned left on the political spectrum, 54\% reported extensive experience with surveys, and 69\% had experience with the World Values Survey (WVS).

\subsection{RQ1: Results}

\begin{figure}[htbp]
    \centering
    \includegraphics[width=\linewidth]{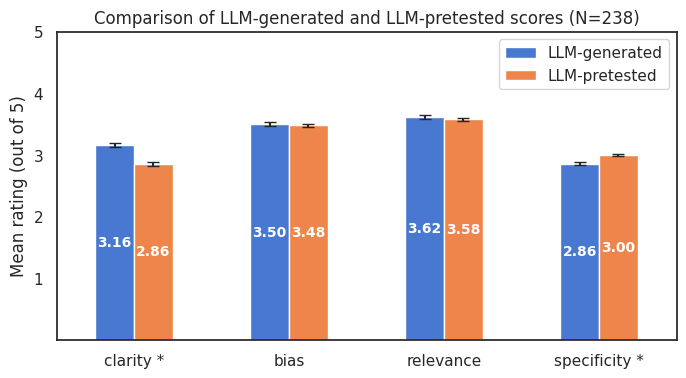}
    \caption{Mean ratings for the LLM-generated and LLM-pretested for the four evaluation criteria (with standard errors) rated by the participants on Prolific for the political trust study. The asterisk ($\ast$) in the label denotes statistical significance ($p<0.05$).}
    \Description{Figure 5 is fully described in the text.}
    \label{fig:political-trust-mean-scores-participants}
\end{figure}

\subsubsection{Prolific participants found LLM-generated statements to be slightly clearer and more specific.}

Figure \ref{fig:political-trust-mean-scores-participants} presents mean ratings on four dimensions—clarity, bias, relevance, and specificity—for LLM-generated and LLM-pretested by prolific users. Participants rated these two conditions, and we report the results.

For clarity, the mean rating for LLM-generated text was 3.16 (SD = 1.13), while for LLM-pretested text, it was 2.86 (SD = 1.10). A t-test indicated a significant difference (p < 0.05), suggesting that the LLM-generated text was perceived as clearer than the LLM-pretested version. For bias, the LLM-generated text received a mean rating of 3.50 (SD = 1.06), while the LLM-pretested text had a mean rating of 3.48 (SD = 1.00). This difference was not statistically significant (p > 0.1), indicating that participants did not perceive one version as more biased. For relevance, the LLM-generated text had a mean rating of 3.62 (SD = 1.01), whereas the LLM-pretested text had a mean rating of 3.58 (SD = 0.98). This difference was also not statistically significant (p > 0.1), suggesting that both versions were perceived as equally relevant. For specificity, the LLM-generated text received a mean rating of 2.86 (SD = 0.83), while the LLM-pretested text was rated higher at 3.00 (SD = 0.81). This difference was statistically significant (p < 0.05), indicating that participants found the LLM-pretested text more specific.

These results suggest a tradeoff between clarity and specificity.  
\textbf{Prolific participants found LLM-generated text was rated as clearer, while LLM-pretested text was seen as more specific.}  
The lack of significant differences in bias and relevance suggests that pretesting did not alter  
how participants perceived these aspects of the text.

\subsubsection{Experts found LLM-generated statements better than LLM-pretested ones across all metrics.}
\label{section:experts-review}

\begin{figure}[!htb]
    \centering
    \includegraphics[width=\linewidth]{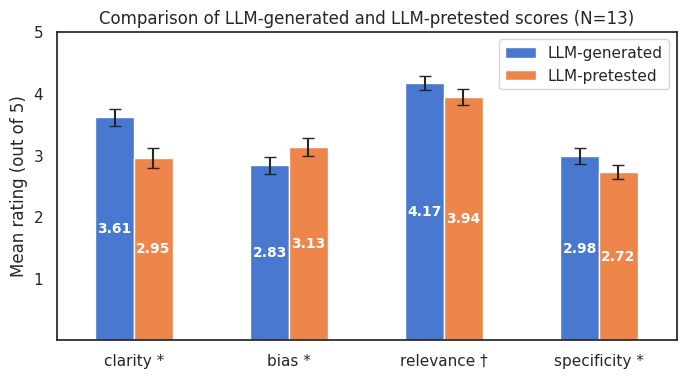}
    \caption{Mean ratings for the LLM-generated and LLM-pretested for the four evaluation criteria (with standard errors) rated by the experts for the political trust study. The asterisk ($\ast$) in the label denotes statistical significance ($p<0.05$) and the dagger ($\dagger$) denotes a marginal statistical significance ($p<0.1$).}
    \Description{Figure 6 is fully described in the text.}
    \label{fig:political-trust-mean-scores-experts}
\end{figure}

Figure  \ref{fig:political-trust-mean-scores-experts} presents the mean ratings on four dimensions—clarity, bias, relevance, and specificity—for LLM-generated and LLM-pretested by experts. Experts rated these two conditions, and we report the findings below.

For clarity, the LLM-generated text received a mean rating of 3.61 (SD = 1.16), while the LLM-pretested text was rated lower at 2.95 (SD = 1.27). A t-test indicated a significant difference (p < 0.05), suggesting that experts found the LLM-generated text significantly clearer than the LLM-pretested version. For bias, the mean rating for LLM-generated text was 2.83 (SD = 1.16), while for LLM-pretested text, it was 3.13 (SD = 1.24). This difference was statistically significant (p < 0.05), indicating that experts perceived the LLM-pretested text as more biased than the LLM-generated version. For relevance, LLM-generated text had a mean rating of 4.17 (SD = 0.88), while the LLM-pretested text had a slightly lower rating of 3.94 (SD = 1.01). This difference was marginally significant (p < 0.1), suggesting that experts found LLM-generated text more relevant. For specificity, the LLM-generated text had a mean rating of 2.98 (SD = 1.07), whereas the LLM-pretested text had a slightly lower mean of 2.72 (SD = 0.94). T-test confirmed a significant difference (p < 0.05), indicating that experts rated the LLM-generated text as more specific than the LLM-pretested text.

Overall, these results suggest that \textbf{experts found LLM-generated text to be significantly clearer, more relevant, and more specific, while also perceiving LLM-pretested text as more biased}. Unlike the findings from the Prolific study, where specificity improved after pretesting, experts rated LLM-generated text higher across most dimensions except for bias, where pretested text was perceived as more biased.

\subsubsection{Participants showed mixed preferences, whereas Experts favored the LLM-generated statements.}

\begin{figure}[!htbp]
    \centering
    \begin{subfigure}[b]{0.9\linewidth}
        \centering
        \includegraphics[width=\linewidth]{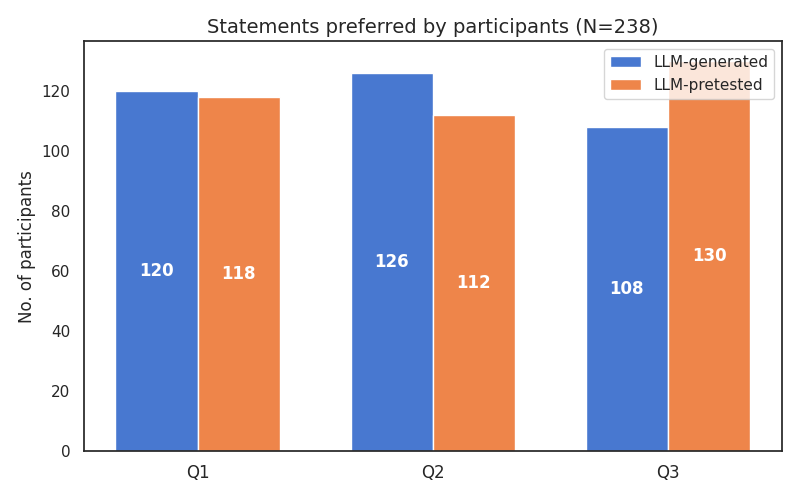}
        \caption{The question preferred by the participants.}
        \label{fig:political-trust-selection-participants}
    \end{subfigure}
    \vfill
    \begin{subfigure}[b]{\linewidth}
        \centering
        \includegraphics[width=0.9\linewidth]{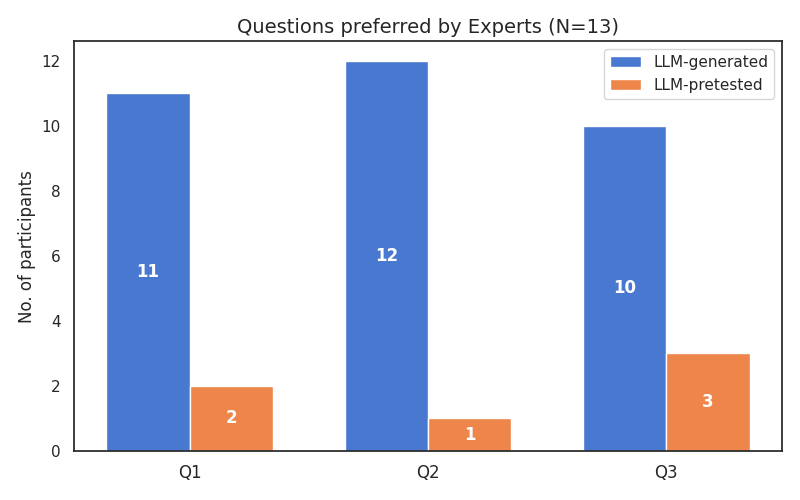}
        \caption{The question preferred by the experts.}
        \label{fig:political-trust-selection-experts}
    \end{subfigure}
    \caption{Comparison of question preferences between participants and experts. Each question (Q1-Q3) includes some statements. Experts evaluated each question with changed and unchanged statements to get the complete context. In contrast, the Prolific participants were shown only the statements that were changed during pretesting to reduce mental load.}
    \Description{The figure shows two bar plots from the question comparison task by the participants and experts respectively in the Political trust study. Each bar shows the number of times a participant or an expert preferred an LLM-generated or LLM-pretested question. The first plot has only slight height differences between the preferences of the LLM-generated and LLM-pretested ones suggesting equal preference to both groups by Prolific participants. The second one shows a large difference in the height of the bars suggesting a clear preference for the LLM-generated questions over the LLM-pretested questions by the experts.}
    \label{fig:political-trust-selection}
\end{figure}

Figure \ref{fig:political-trust-selection} presents the distribution of question preferences for LLM-generated and LLM-pretested statements among participants (Figure \ref{fig:political-trust-selection-participants}  top) and experts (Figure \ref{fig:political-trust-selection-experts} bottom). Among participants (N=238), preferences were relatively balanced across conditions, with no strong preference for either LLM-generated or LLM-pretested questions. However, among experts (N=13), there was a clear preference for LLM-generated statements, with a majority selecting them over the pretested versions for all three questions. This contrast suggests that while general participants found both versions similarly acceptable, experts perceived the LLM-generated statements as preferable. These findings align with previous results indicating that experts rated LLM-generated content as clearer and more relevant than LLM-pretested versions.

From open-ended responses, prolific participants preferred clarity and simplicity over excessive detail, emphasizing neutrality and relevance in survey statements. Participants overwhelmingly favored LLM-generated statements for their concise and direct phrasing, with one noting, \textit{LLM-generated is preferable because it is more concise. Although LLM pre-tested is a better question, it is not as clear, and too many similar type questions may lead the participant to become fatigued.} However, some responses indicated that LLM pre-tested statements offered more precision, as one participant pointed out, \textit{LLM-generated questions are far less specific than LLM pre-tested questions.} Bias emerged as a concern, particularly with leading or overly complex phrasing. One participant noted, \textit{Both statements in LLM pre-tested are too convoluted regarding the adjectives used to describe various nouns, making the statements a little leading.} Relevance also played a role, particularly when discussing government actions. Participants preferred specific references over general statements, as one participant stated, \textit{It mentions the government’s record. LLM-generated just says actions, while LLM pre-tested gives specifics.} Overall, participants valued clarity and neutrality while recognizing the trade-off between simplicity and detail in survey design.

From open-ended responses, experts preferred clarity and simplicity over excessive specificity, highlighting the trade-off between readability and precision in survey design. Experts generally favored LLM-generated  statements for their shorter, more understandable phrasing, with one noting, \textit{"Shorter sentences -> more understandable,"} and another emphasizing, \textit{"Simple and straightforward statements."} Readability was a key concern, particularly for a general audience, as one expert pointed out, \textit{"The reading age for Group 2 (LLM pre-tested) questions is too high. Also, all the Group 2 green questions ask about two things simultaneously, so aren't clear"}. While LLM pre-tested statements were recognized as more precise, some experts found them too complex or difficult to interpret. One expert stated, \textit{"Although the statements in Group 2 seem more precise, they are more difficult to answer for the general public because they are too specific (for example, `public welfare and policy outcomes' instead of `interests of the public')."} Another highlighted a trade-off, noting that specifics are reasonable, but complex wording and excessive detail made certain statements unnecessarily complicated. Overall, experts valued clarity and accessibility while acknowledging the importance of specificity when it did not hinder comprehension.

\section{RQ2: Adapting questionnaires for cross-cultural surveys}
\label{section:adapting-sa}

In this experiment, we looked at the potential of Large Language Models in adapting existing questionnaires for a specific target audience. We were interested in determining if a questionnaire on climate change tested in the United States could be adapted to an African country using our pretesting pipeline with LLMs. We decided on South Africa as the target audience due to the high number of available users on Prolific.

\subsection{RQ2: LLM adapted US-specific questionnaire for South African audience.}
\subsubsection{RQ2: Setup}
We obtained a standardized survey from a study focused on U.S. state-level public opinion about climate change from 2008 to 2020\cite{Marlon2022}. Thirty questions assess public perceptions of global warming, including beliefs about its existence, human causes, potential harm, and policy support.\footnote{\url{https://climatecommunication.yale.edu/visualizations-data/ycom-us/}} The questions were extracted and sanitized, i.e., correctly formatting for the LLM, e.g., replacing the United States with South Africa. Then, we used an LLM to summarize the questionnaire as a research question. Lastly, using the questionnaire generation pipeline (section \ref{ref:questionnaire-generation}), we generated a list of South African participants’ personas based on the research question. We disregarded the generated questionnaire because we already had the standardized version, and our focus in this experiment was on the generated personas. We then used the list of personas in our pretesting pipeline to evaluate the US questionnaire and determine necessary changes for a South African audience. All 30 questions were pretested, but the LLM suggested changes to only five questions, some of which are shown in Table \ref{tab:rq2-questions-list}.

\begin{table*}[ht]
    \centering
    \begin{tabular}{|p{5cm}|p{5cm}|p{7cm}|}
       \hline
       \textbf{\textcolor{LLM-generated}{Question from the standard survey}}  & \textbf{\textcolor{LLM-pretested}{Question after LLM-pretesting}} & \textbf{LLM reasoning} \\
       \hline
       Assuming global warming is happening, do you think it is...? \newline Options: Caused mostly by human activities, Caused mostly by natural changes in the environment, None of the above because global warming isn’t happening, Other, Don’t know  & Assuming global warming is happening, what do you believe is the primary cause?
       \newline Options: Caused mostly by human activities, Caused mostly by natural changes in the environment, Don't know, Other (please specify) & This question can be problematic because it includes an option ("None of the above because global warming isn’t happening") that contradicts the assumption made in the question. This might confuse the respondents or lead to inconsistencies in their answers. It's important that the question aligns with the premise. Also, the option "Other" seems vague, which can lead to confusion about what respondents should include under this category. \\
       \hline
       How often do you discuss global warming with your friends and family? \newline Options: Often, Occasionally, Rarely, Never & How often do you have in-depth discussions about global warming with your friends and family? \newline Options: Often, Occasionally, Rarely, Never & The question lacks a clear delineation of what constitutes a discussion about global warming. This can result in variation among participants in terms of what intensity or depth qualifies as a discussion. \\
       \hline
    \end{tabular}
    \caption{The original questions, the questions after making the changes suggested by the LLM, and the reasoning given by the LLM for the changes. The remaining questions are shown in the Appendix in Table \ref{tab:rq2-questions-list-remaining}.}
    \label{tab:rq2-questions-list}
\end{table*}

\subsubsection{RQ2: Participants}
We sampled participants aged 18 and above from South Africa who had at least an undergraduate degree in subjects such as Earth, Environment or Climate Sciences, Politics, Psychology, Social Work, or Sociology. A total of 118 participants participated in the study, 75\% of which identified as female. About 31\% of the participants defined themselves as leaning toward the political center-left, and 99\% mentioned having at least some experience with surveys.

\begin{figure}[H]
    \centering
    \includegraphics[width=\linewidth]{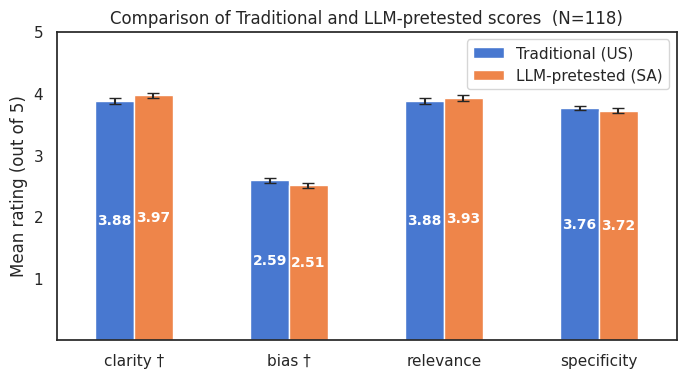}
    \caption{Mean ratings (with standard errors) for the original US-based (traditional) questionnaire and the questionnaire adapted for the South African (SA) audience after LLM-based pretesting. The dagger ($\dagger$) symbol denotes a marginal statistical significance ($p<0.1$).}
    \Description{Figure 8 is fully described in the text.}
    \label{fig:cross-cultural-mean-ratings}
\end{figure}

\subsection{RQ2: Results}

Figure \ref{fig:cross-cultural-mean-ratings} presents ratings on four dimensions—clarity, bias, relevance, and specificity. Participants rated the traditional standardized and LLM-pretested content on our scale above, and we analyzed their differences.

For clarity, the traditional questions received a mean rating of 3.87 (SD = 1.12), while the pretested questions were rated slightly higher at 3.97 (SD = 1.07). A t-test indicated a marginally significant difference (p < 0.1), suggesting that participants found the pretested questions somewhat clearer than the traditional ones. For bias, the traditional questions had a mean rating of 2.58 (SD = 1.01), while the pretested questions were rated slightly lower at 2.51 (SD = 1.03). This difference was also marginally significant (p < 0.1), indicating that pretesting slightly reduced perceived bias in the questions. For relevance, the traditional questions had a mean rating of 3.88 (SD = 1.13), while the pretested questions received a similar rating of 3.93 (SD = 1.11). This difference was not statistically significant (p > 0.1), suggesting that pretesting did not notably impact how relevant participants found the questions. For specificity, the traditional questions received a mean rating of 3.77 (SD = 0.87), whereas the pretested questions were rated slightly lower at 3.72 (SD = 0.93). A t-test confirmed that this difference was not significant (p > 0.1), indicating that participants found both traditional and pretested questions similarly specific.

Overall, these results suggest that \textbf{Prolific participants found LLM-adapted questions slightly clearer and less biased than traditional questions, though differences were minor.} No significant changes were observed for relevance or specificity, indicating that pretesting had minimal impact on these aspects.

\begin{figure}[htb]
    \centering
    \includegraphics[width=\linewidth]{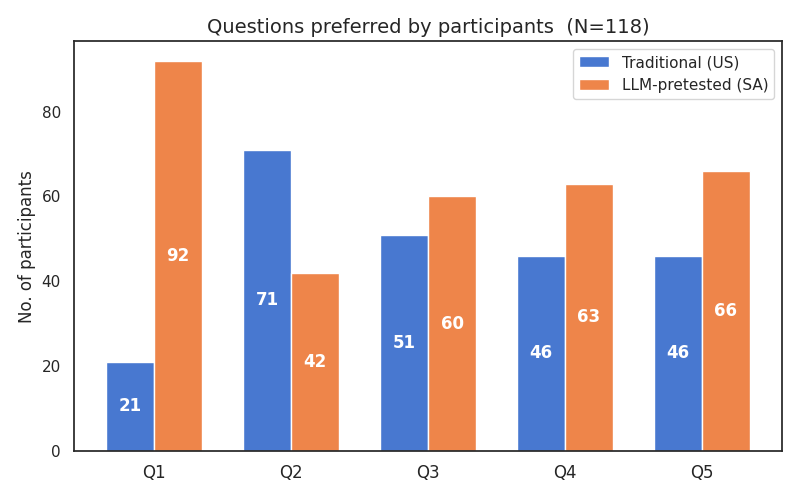}
    \caption{The results of the questionnaire comparison task where the participants were given the original and pretested questions side-by-side and asked to select the question they thought was clearer. We can see that in 4 out of the 5 pairs of questions tested, the participants selected the pretested question as being clearer and easier to understand.}
    \Description{The figure shows a bar plot with 5 pairs of bars one for each pair of LLM-generated and LLM-pretested questions tested in the questions comparison task in the Questionnaire adaptation study. In 4 out of the 5 questions, participants show a clear preference for the LLM-pretested questions.}
    \label{fig:cross-cultural-selection}
\end{figure}

For the comparison task, participants picked the pre-tested question for 4 out of 5 in this task, as shown in Figure \ref{fig:cross-cultural-selection}. From open-ended responses, participants preferred LLM pre-tested to the traditional for its clarity, specificity, and conciseness, emphasizing the need for precise yet accessible wording in survey design. Participants found LLM-pretested clear as it conveyed the intended meaning transparently. One participant stated \textit{"LLM-pretested is clearer because it is more comprehensive and detailed,"} and another said, \textit{"Straight to the point, very clear."} Specificity was another key factor, as LLM-pretested made answering questions easier. A participant pointed out, \textit{".. asking for a specific answer in terms of contextualizing the extent to which one discusses global warming"}. However, a few participants raised concerns about bias in phrasing, noting that terms like \textit{"in-depth"} (Q2) could introduce unintended assumptions, with one stating, \textit{"Because Q2 is biased by the word 'in-depth'"}. Participants also preferred simple language, which helped them process easier, \textit{"Because it is written in simple English, no big words."} Overall, participants valued clear, specific, and concise language while being mindful of potential bias in question phrasing.

\section{Discussion}
In this work, we explored the potential of Large Language Models in generating and pretesting survey questionnaires. Compared to existing research, we focused primarily on generating specific questions related to the research question in the questionnaire and using a simulated pilot study to pretest the generated questionnaire. Furthermore, we conducted multiple Prolific studies to evaluate the LLM-generated and LLM-pretested content. Below, we reflect on the results and what they mean regarding the use of LLMs in survey research.

\subsection{Empirical Insights on the Suitability and Challenges of LLMs in Survey Design}

\subsubsection{RQ1: LLM-generated questions were more favorable than LLM-pretested for experts and Prolific participants.}

In our first experiment (see Section \ref{section:political-trust}), we compared LLM-generated statements with LLM-pretested statements. Building on prior work \cite{lei_qsnail_2024}, we used two additional evaluation metrics—bias and clarity—assessed by human evaluators. From the mean rating scores presented in Figure \ref{fig:political-trust-mean-scores-participants} (238 Prolific participants) and Figure \ref{fig:political-trust-mean-scores-experts} (13 experts), we found that LLM-generated statements were clear, slightly biased, and highly relevant compared to the LLM-pretested ones. Overall, both experts and participants favored LLM-generated over LLM-pretested questionnaires. Open-ended responses from experts revealed that pretesting made the questions more wordy and complex which can affect comprehensibility and data quality as suggested by ~\citet{lenzner_effects_2012}. As a result, in this scenario, pretesting did not improve the generated questionnaires. In fact, in one case, expert response suggested that pretesting even affected the meaning of the statements (\textit{Questions that I would answer with yes in LLM-generated would be `no' in LLM-pretested}). We suspect that the vast training corpus of LLMs, which likely includes standardized questionnaires~\cite{rothschild_opportunities_2024}, enables them to generate high-quality questions without additional refinement through pretesting. Therefore, future research can utilize our pipeline by examining various topics, domains, and models to assess the pipeline's suitability for particular domains or topics.


\subsubsection{RQ2: LLM-adapted questions were more favorable than traditional ones for the South African audience.}
In our second experiment (see Section \ref{section:adapting-sa}), we evaluated our LLM-based pretesting pipeline to adapt a U.S.-based questionnaire for a South African audience. Figure \ref{fig:cross-cultural-mean-ratings} presents the mean ratings for both the standardized and pretested questionnaires. Participants generally found the original U.S.-based questionnaire clear, relevant, slightly biased, and specific. Pretesting led to slight improvements in clarity and reduced bias, indicating that participants perceived the pretested questionnaire as better suited for a South African audience. In the comparison task, 4 out of 5 LLM-pretested questions (figure \ref{fig:cross-cultural-selection}) were preferred over the traditional questionnaire, as the longer pretested questions provided better content clarification. However, specific keywords added by the LLM, such as “in-depth” (1/5), reduced clarity or introduced confusion. Participants also noted that longer wording made the questions harder to read. Our findings suggest that LLMs can enhance survey design when adapting questionnaires across geographic regions. Future research could explore how culture, language, and regional differences influence survey adaptation and effectiveness.

\subsection{Pipeline for using LLMs for Generation, Pre-testing, and Adaptation.}

We developed an innovative pipeline that uses LLMs to support the survey questionnaire design process. Building upon the works by \citet{lei_qsnail_2024, laraspata_enhancing_nodate, rothschild_opportunities_2024}, we expanded their work using Retrieval Augmented Generation (RAG) to enhance questionnaire generation. Given a research question, our pipeline generates questionnaires by incorporating existing standardized questions and relevant news articles. Additionally, it creates a set of fictitious personas that can be used to simulate a pilot study and provide feedback on the generated questions, potentially improving the questions. Our pipeline also enables the adaptation of existing questionnaires to different contexts, such as tailoring a survey initially designed for the U.S. to suit the South African public. We developed and deployed human evaluation metrics to assess the quality of questions generated by our pipeline. Our evaluation is based on four key criteria: relevance and specificity (adapted from ~\citet{lei_qsnail_2024}), along with two widely used criteria, clarity, and bias. To enhance reliability, we introduced a paired evaluation approach, where each criterion is assessed directly and indirectly. For transparency and reproducibility, we have shared our prompts, LLM-generated feedback, and evaluation questions (see the Appendix). Additionally, our code is available publicly for researchers to use and improve upon\footnote{https://github.com/Societal-Computing/automated-questionnaire-pretesting}. Our pipeline is flexible and, thus, can be easily extended or combined with existing LLMs and applied to new domains. Our pipeline could benefit survey researchers seeking quick feedback on their questionnaire drafts, thereby reducing the need for multiple human pretesting iterations. We expect our pipeline and evaluation metrics can contribute to advancing the relatively sparse research on LLM-enhanced survey design.

\subsection{Better Questionnaire design: Balancing Clarity, Specificity, and Neutrality?}

Through open-ended responses from experts and participants, we found that effective questionnaires must balance clarity, specificity, and neutrality to improve data collection. Clarity is essential, as overly complex wording can reduce readability and increase cognitive burden~\cite{kamp_using_2018}. At the same time, specificity is crucial—questions that are too brief, overly simplistic, or filled with jargon may lack precision, ultimately affecting readability and comprehension~\cite{taherdoost_designing_2022}. This highlights the need to manage the tradeoff between readability and specificity carefully. Additionally, bias can be introduced through vague word choices, such as “in-depth,” which may unintentionally influence responses~\cite{lenzner_effects_2012}. This underscores the importance of using neutral language to maintain objectivity in survey design. Participant feedback is key in refining our pipeline and generating survey questions that better align with new target audiences. Future work could extend our pipeline by integrating open-ended input from users, enabling iterative improvements—such as refining prompts based on human participant feedback—to enhance question quality and relevance further.

\section{Limitations and Future work}
While our analysis shows the potential of LLMs in survey research, there are limitations to the approach we used in our experimentations. First, we only focused on generating simple questionnaires when existing questionnaires often have complex flow and branching logic. We generated closed-ended and open-ended questions but didn't include other types of questions, such as matrix-style questions. We also didn't include demographics-related questions in the generated questionnaires. Furthermore, we didn't evaluate the overall quality of the questionnaire (e.g., coherence, logical flow, etc). Pretesting is inherently an iterative process; however, in our experiments, we limit our analysis to the initial iteration and do not implement subsequent rounds using the revised questionnaire. Establishing a definitive stopping criterion for such iterations proves challenging, as contemporary large language models (LLMs) tend to consistently comply with prompts and continue suggesting modifications in each cycle. However, LLM-generated content may exert an anchoring effect, particularly on individuals from the general public with limited experience \cite{rothschild_opportunities_2024}, as the output can appear highly plausible. Therefore, we recommend that our proposed pipeline be used as a complementary tool within established survey design processes for researchers, and as an educational resource for those learning about survey methodology.

Similarly, in the study design, we selected participants with relevant domain knowledge and survey experience, providing them with detailed explanations of the four evaluation criteria, illustrative examples, and test questions to assess their understanding. Nevertheless, we cannot be certain that participants fully internalized the criteria or applied them as intended during their evaluations. For instance, in the question preference task within the Political Trust study, many participants appeared to base their preferences on personal opinions about the content of the statements (`\textit{I selected that because the government clearly does not fully execute all their promises.}') rather than on an objective assessment of sentence structure and clarity. Likewise, multiple elements, including the duration of the study and participant biases, can influence self-reported measurements such as those in our studies. Assessing these questions is cognitively demanding, and participants may hasten through the latter sections of the study, particularly the question preference and open-response tasks. This could impact the reliability of self-reported evaluations. Consequently, researchers should exercise caution when employing crowdsourced participants from platforms such as Prolific. Ideally, a more representative sample should be used, and participants' overarching ideological orientations and potential biases should be assessed before the evaluation of survey items. Moreover, the sample size of experts (13) was small compared to participants on Prolific and while the difference in preference was more pronounced, the results might not be replicable with a larger sample size. Such a small sample size could compromise the statistical significance and increase the risk of sampling bias.

Additionally, LLMs have biases based on their training and finetuning data, which could affect the generated questionnaires~\cite{https://doi.org/10.1002/pra2.1061, Yang2024, Hofmann2024, gupta_bias_2024} in various domains. For example, LLMs are shown to have a center-left leaning~\cite{10.1371/journal.pone.0306621}, which could affect questionnaire generation and pretesting in the Political domain. Similarly, LLMs can reinforce stereotypes, and misrepresent underrepresented groups~\cite{wang_large_2024}. Moreover, in the context of low-resource languages, LLMs tend to exhibit reduced performance, primarily due to the limited availability of training data~\cite{dargis-etal-2024-evaluating}. Thus, further evaluation of the generated questionnaires should be done to determine such bias seeping into the generated questions.

Future research could expand upon our work to create complex-structured questionnaires with diverse question types along with specific demographic questions tailored to the research question, enabling researchers to gather targeted information about their audience. Depending on the research question, questionnaires could incorporate images or even video media. With the rising popularity of Vision-Language-based LLMs, developing multi-modal survey instruments could be a promising goal. Further refining the questionnaire generation prompt with more examples could enhance results by specifying how to effectively utilize the provided context, such as determining the number of questions to generate from the given news summary, determining the number of questions to reuse or modify from existing questionnaires, and determining the amount of focus to put on the research question. The feedback from experts in our study provided valuable criteria for designing questions, which could inform better questionnaire generation prompts. Fine-tuning LLMs with existing pretesting reports could help develop LLMs that are better capable of detecting issues with the generated questionnaires and offering better suggestions. Future work could explore designing interviewer agents capable of implementing Cognitive Interviewing techniques effectively in Pilot study simulations. Creating LLM personas that respond diversely, like humans, could make pretesting results more realistic. In our pipeline, we provided the reviewer LLM only with the interview transcripts to identify questionnaire issues. This could be supplemented by using the reviewer for behavioral coding and annotating the interview transcripts for further analysis. Similarly, including more examples in the reviewer's prompt could make the recommendations more specific. Furthermore, our work can be expanded to low-resource languages or other domains by using LLMs trained in specific languages or domains that could be particularly effective in generating relevant questions or simulating participants from those contexts, helping to address cultural nuances that commercial LLMs may overlook.

\section{Conclusion}
In conclusion, our work demonstrated the potential of Large Language Models (LLMs) in enhancing the design and pretesting of survey questionnaires. Our work contributed empirical evidence from two Prolific studies, highlighting both the strengths and challenges of using LLMs in survey design. We also introduced a novel LLM-based pipeline for automated questionnaire generation and pretesting, which can be adapted across various domains and LLMs. Our findings indicate that LLMs, such as GPT-4o, with relevant contextual information, can effectively generate relevant and specific questions, reducing the need for extensive human pretesting. However, while LLMs showed promise in creating high-quality questions, their ability to pretest survey questionnaires remained limited, suggesting that human oversight is still necessary to ensure question clarity and relevance. For adapting existing questionnaires to specific target audiences, our results revealed that LLMs can make questionnaires more explicit, unbiased, and relevant. This capability is particularly valuable in cross-cultural studies, where nuanced adaptations are crucial for accurate data collection.

While LLMs offer significant advantages, it is essential to acknowledge their limitations. Future research should focus on refining LLM capabilities to better handle the subtleties of questionnaire design and pretesting, ensuring that surveys remain a reliable tool for gathering meaningful insights across diverse populations.

\begin{acks}
We want to acknowledge the members of the Interdisciplinary Institute for Societal Computing at Saarland University for helping with the design and testing of the study experiments. Divya Mani Adhikari, Vikram Kamath Cannanure, and Ingmar Weber are supported by funding from the Alexander von Humboldt Foundation and its founder, the German Federal Ministry of Education and Research.
\end{acks}

\bibliographystyle{ACM-Reference-Format}
\bibliography{references}

\appendix

\onecolumn

\section{Prompts}
\label{appendix:prompts}

\subsection{Questionnaire generation system prompt}
\noindent\fbox{\begin{minipage}{\textwidth}
{\fontfamily{cmtt}\selectfont\small
You get a research question describing the hypothesis the researcher wants to test along with a summary from relevant recent news article as extra context or a list of relevant questions from existing standardized surveys. 
Based on that you devise a survey questionnaire.  Make sure to include at least one question each from the given news summary and list of relevant questions but do not mention the source in the question.
Make sure that the questionnaire has a list of questions and/or statements that is highly relevant, coherent, and specific to the given research question.  
\\
Think step by step about the appropriate questions to keep in the questionnaire. 
Each question should be particular to the given research question. 
Ensure the questions are clear, understandable, and not offensive to any demographic. 
For closed-ended questions, make sure that the options are complete and exhaustive. 
Ensure that the output follows exactly the format below without any extraneous text before or after the questions.
If the instructions mention a specific language, generate a survey using that language. Otherwise, use English.
\\
\\
Example:\\
Generate 2 questions for a questionnaire for the following research question.\\
\\
\textless question\textgreater Why does the public like this product?\textless /question\textgreater\\
\\
\textless relevant\_articles\textgreater\\
1. Summary of the first article\\
\textless/relevant\_articles\textgreater\\
\\
\textless relevant\_questions\textgreater\\
* First relevant question\\
\textless /relevant\_questions\textgreater\\
\\
Questions:\\
1. Why do you like this product?\\
   Type: closed-ended\\
   Options: Ease of use, Quality, Price, Brand\\
\\
2. What suggestions do you have for the product?\\
   Type: open-ended\\
   Options: -
}
\end{minipage}}

\subsection{Persona generation system prompt}
\noindent\fbox{\begin{minipage}{\textwidth}
{\fontfamily{cmtt}\selectfont\small
You are a researcher working on Survey research. As a part of your research, you need to determine the correct samples to pick from the intended audience of the survey. You will be given a research question. Based on that, you need to create a list of personas that represent the ideal participant for the survey. \\
Generate just the personas without any extraneous texts or markdown syntax.\\
The  persona should  contain at least these basic information:\\
- Name (This should be a fictional name that is common in the intended demographic of the survey)\\
- Age\\
- Gender\\
- Race\\
- Location\\
- Occupation\\
- Education\\
- Preferences (based on the research question and other context information)\\
You can also add more information to the persona if you think it is necessary (such as ethnicity, specific preferences, etc). Make sure to create a varied list of personas to cover all the possible demographics of the intended audience. Think carefully while generating the list of personas and be specific with the details.\\
Example:\\
Generate 10 personas for the following research question:\\
\textless question\textgreater What is the food preference of young adults in America?\textless/question\textgreater \\

Persona 1:\\
Name: John Doe\\
Age: 22\\
Gender: Male\\
Race: White\\
Location: California, USA\\
Education: Bachelors in Computer Science\\
Preferences: Vegan diet\\
}
\end{minipage}}

\subsection{System prompt to detect Language \& Domain to retrieve relevant questions from SQP}
\noindent\fbox{\begin{minipage}{\textwidth}
{\fontfamily{cmtt}\selectfont\small
You are given a research question based on which you will define the parameters for a function call. You return the parameters in JSON format. The details of the parameters are as follows:\\

* Language\\
   Type: String\\
   Options:  `German', `French', `Dutch', `Czech', `Danish', `Swedish', `Finnish', `Greek', `Hungarian', `English', `Hebrew', `Arabic',  `Russian', `Italian', `Norwegian', `Polish', `Portuguese', `Slovene', `Spanish', `Catalan', `Estonian', `Icelandic', `Slovak', `Turkish', `Ukrainian', `Bulgarian', `Latvian', `Romanian', `Croatian', `Lithuanian', `Albanian'
   \\
   Note: If the required option isn't available, select 'English' as the option.\\

* Domain\\
   Type: String\\
   Options: `Leisure activities', `Other domains', `National politics',  `Living conditions and background variables', `International politics', `Family', `Personal relations', `Work', 'Health', `European Union politics', `Consumer behaviour'\\

Example:
Generate the parameters for the following research question:
\\
\textless research\_question\textgreater
\\
What is the perspective of people in Germany about European politics?
\\
\textless /research\_question\textgreater\\

Output:\\
\{"language": "German", "domain": "European Union politics"\}
}
\end{minipage}}

\subsection{System prompt to summarize relevant News}
\noindent\fbox{\begin{minipage}{\textwidth}
{\fontfamily{cmtt}\selectfont\small
You will be provided excerpts from 5 news articles. Summarize each article in 1 sentence i.e. 5 sentences in total.
}
\end{minipage}}

\subsection{Follow-up Question generation system prompt}
\noindent\fbox{\begin{minipage}{\textwidth}
{\fontfamily{cmtt}\selectfont
You will be given a chat history with a participant and you generate a probing question to probe the participant.
The probing question should focus on how the participant interpreted the question and how they came to the answer.

Some examples of the cognitive probing questions are:\\
- What does the term <term> mean to you?\\
- How did you come up with your answer?\\
- Can you repeat the question I just asked you in your own words?\\
- How sure are you that <answer>?\\
- How do you remember that you have <answer>?\\
- Why do you say that you think it is <answer>?\\
- Was that easy or hard to answer?\\
- I see you have <answer>. Can you tell me more about that?\\
- What do you think the purpose of this question is?\\
- I noticed that you hesitated before answering. Tell me what you were thinking.\\
- Tell me more about that.\\

Use above template along with the chat history to generate specific probing questions.
Detect the appropriate language to use based on the context of the chat history. Otherwise, use English.
The output should be just the question without any numbering or additional text. 
}
\end{minipage}}

\subsection{(Example) Participant prompt}
\noindent\fbox{\begin{minipage}{\textwidth}
{\fontfamily{cmtt}\selectfont\small
You are a participant in a survey interview for a study. Following are your details:\\

Name: Thabo Dlamini \\
Age: 35 \\
Gender: Male \\
Race: Black \\
Location: Johannesburg, South Africa \\
Occupation: Environmental Scientist \\
Education: Masters in Environmental Science \\
Beliefs: Strong believer in climate change and its impacts \\
Perceptions of Risk: High concern about air pollution and water scarcity \\
Policy Preferences: Supports government regulations on carbon emissions \\
Behaviors: Actively recycles and uses public transportation \\

You will be asked a series of questions based on the survey questionnaire. 
Please answer them strictly following the details above and the chat history so far below. 
If the question has options strictly pick just the option. For open-ended questions, answer in maximum 2-3 sentences.
If the question is in a language other than English, please answer in that respective language.
Please stick strictly to the details provided above and do not deviate from them.
}
\end{minipage}}

\subsection{Reviewer prompt}
\noindent\fbox{\begin{minipage}{\textwidth}
{\fontfamily{cmtt}\selectfont\small
As a reviewer, you have been provided with the transcripts of a survey pilot study interviews, along with the original research question. 
Your task is to evaluate the survey questionnaire based on several criteria, including clarity, comprehension, and sensitivity. 
Select a question from the interview transcript where the participant encountered difficulties or misunderstandings.
Check for any contradictions or inconsistencies in the participant's responses. \\

Some of the possible issues that you need to look for are: \\
double questions\\
ambiguous questions\\
ambiguous word meanings\\
loaded or leading questions or phrase\\
level of question difficulty\\
lopsided response categories\\
missing response categories\\
missing questions\\
necessity and relevance of individual questions\\
discriminating questions (between certain groups within the target group)\\
non-response rates\\
effect of ordinal position of multiple responses\\
perceptions of pictures\\
degree of attention\\

Provide a detailed review of the survey questionnaire, highlighting any areas that need improvement or revision as shown in following example. 
Then, suggest the corrected questions as well.
Consider only the main questions when suggesting improvements and ignore follow up questions when suggesting corrections. 
Think carefully about the participant's responses to the questions and how they could be improved.
When describing issues, mention the question as well as the participant number (e.g. Persona 1) and what problem you found with it.

Example output:\\

Question: Question from the survey questionnaire.\\
Problem: Describe the problem with the question. Relate it to the above points.\\
Sources: Mention the participant number and the transcript where the issue was observed.\\
Suggestion: Suggest a corrected version of the question.\\

Question: Question from the survey questionnaire.\\
Problem: Describe the problem with the question. Relate it to the above points.\\
Sources: Mention the participant number and the transcript where the issue was observed.\\
Suggestion: Suggest a corrected version of the question.\\
}
\end{minipage}}

\subsection{Political trust Questionnaire prompt}
\noindent\fbox{\begin{minipage}{\textwidth}
{\fontfamily{cmtt}\selectfont\small
Generate a questionnaire for a survey on political trust in the United States. The questionnaire should evaluate political trust as a latent concept so instead of directly asking the participants, the questionnaire should measure political trust using other aspects based on the following theory.\\

Beliefs are shaped by a mix of cognitive and affective processes, adapting to new information or stabilizing as enduring political attitudes. Cognitive processes involve using knowledge and experience to form expectations about political actors or institutions based on benevolence (caring for others), integrity (keeping promises), and competence (ability to deliver). Affective processes are driven by emotions tied to shared values, norms, and group identities. These influences guide "intentions to act," where individuals take actions that demonstrate trust by making themselves vulnerable to political actors or institutions. Cognitive, affective, and intention-based trust interact dynamically, with their relative impact varying by individual and context.\\
\\
There are different sections in the questionnaire. Generate each section based on the following instructions.\\

1. Generate 6 statements to rate on a likert scale (1-5, 1 meaning Agree strongly and 5 meaning disagree strongly) on the question "How you feel about the national congress in your country?". Number it from P1 to P6.\\
2. Generate 6 statements to rate on a likert scale (1-5, 1 meaning Agree strongly and 5 meaning disagree strongly) on the question "How do you feel about the government in your country?". Number it from G1 to G6.\\
3. Generate 6 statements to rate on a likert scale (1-5, 1 meaning Agree strongly and 5 meaning disagree strongly) on the question "How do you feel about the United Nations?". Number it from UN1 to UN6.\\
4. Generate 15 statements to rate on a likert scale (1-5, 1 meaning Agree strongly and 5 meaning disagree strongly) related to trust on Politicians and the government. Number it from A to O.\\
5. Generate a question asking about their trust on the \textless head of state\textgreater in their country. Replace \textless head of state\textgreater with the respective head of state in the country. Make sure to use a likert scale of 0-10, 0 meaning no trust at all and 10 meaning a great deal of trust.\\
}
\end{minipage}}

\clearpage
\section{Prolific Study design}
\label{appendix:prolific-study-design}
\begin{figure}[htb!]
    \centering
    \includegraphics[width=0.5\linewidth]{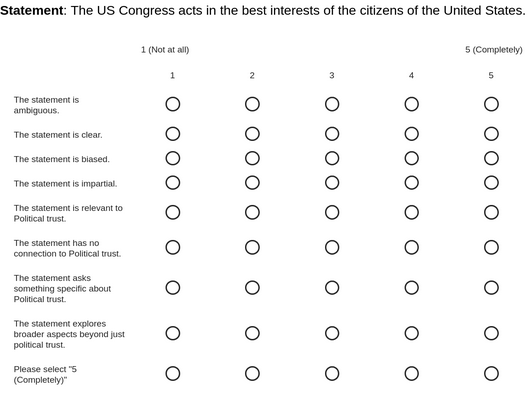}
    \caption{Example question from rating task of the study where participants were asked to evaluate the given statement on a Likert scale.}
    \Description{The figure shows a screenshot of one of the statements from the Political trust study used in the question rating task. The statement to be rated is on the top and below it are the 8 evaluation statements used to evaluate the questions in our four criteria using a 5-point Likert scale (1 being 'Not at all' and 5 being 'Completely') along with an attention check statement where participants have to select a specific value on the Likert scale.}
    \label{fig:political-task-1}
\end{figure}

\begin{figure}[htb!]
    \centering
    \includegraphics[width=0.5\linewidth]{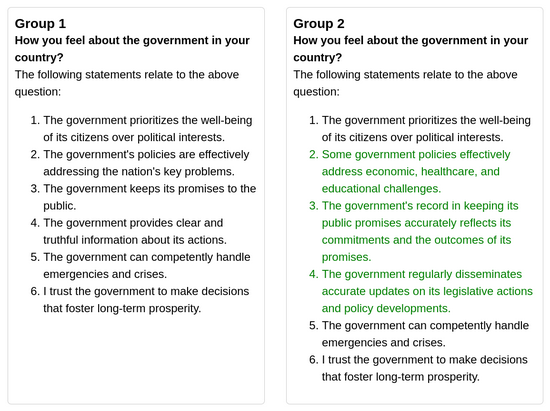}
    \caption{Example question from comparison task of the study where participants were asked to select the clearer question.}
    \Description{The figure shows a screenshot from the question comparison task from the Political trust study. It shows two boxes side by side titled Group 1 and Group 2. Group 1 is the LLM-generated question and Group 2 is the same question with the changes suggested by LLM-pretesting. The parts changed after LLM pretesting have been highlighted in green.}
    \label{fig:political-task-2}
\end{figure}
\newpage
\section{RQ1: Political trust study}
\label{appendix:political-trust-study}

\begin{figure}[htb!]
    \centering
    \begin{subfigure}{0.45\textwidth}
        \includegraphics[width=\textwidth]{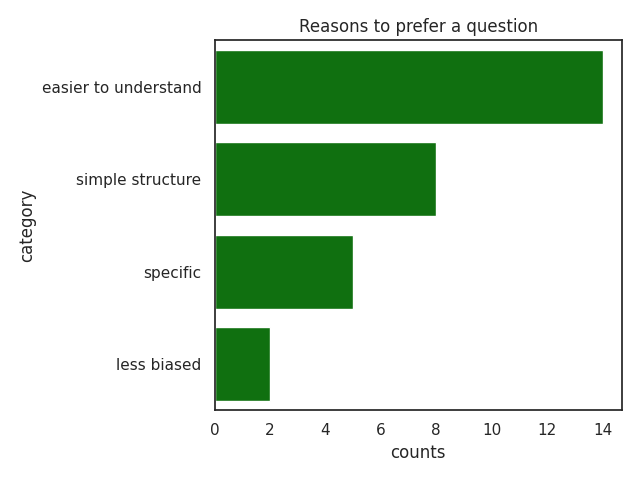}
    \end{subfigure}
    \hfill
    \begin{subfigure}{0.45\textwidth}
        \includegraphics[width=\textwidth]{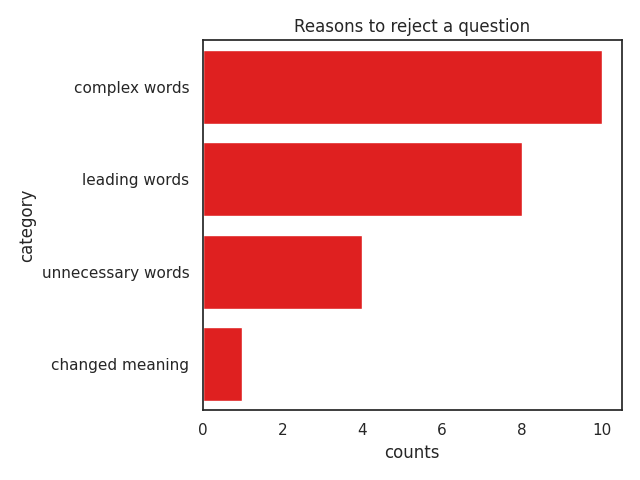}
    \end{subfigure}
    \caption{Manually coded reasons from the free text responses of the experts when selecting the clearer question in the question comparison task.}
    \Description{The figure shows two horizontal bar plots from the manual annotation of the free text responses of the experts in the Political trust study. The left plot shows the reasons the experts mentioned while preferring a question. The most common reason is that the question was 'easier to understand' followed by 'simple structure', 'specific', and 'less biased'. The second plot shows the reasons experts mentioned when disliking a question. The most common reason is that the question had 'complex words', followed by 'leading words', 'unnecessary words', and 'changed meaning'.}
    \label{fig:experts-manual-coding}
\end{figure}

\begin{table}[htb!]
\begin{tabular}{|p{4cm}|p{5cm}|p{8cm}|}
       \hline
       \textbf{Statement generated by LLM}  & \textbf{Statement after LLM-pretesting} & \textbf{LLM reasoning} \\
        \hline
        The government keeps its promises to the public. & The government's record in keeping its public promises accurately reflects its commitments and the outcomes of its promises. & Participant 20 (Priya Nair) indicated a "Disagree" response, but mentioned they hadn't yet been asked that question, suggesting confusion with the survey's flow or misunderstanding of the statement's meaning.\\
        \hline
        The government provides clear and truthful information about its actions. & The government regularly disseminates accurate updates on its legislative actions and policy developments. & The phrase "clear and truthful information" is subjective, leading to varied interpretations by participants, which affects the ability to respond accurately about the government's communication practices.\\
        \hline
        Politicians prioritize the interests of the public over their own. & Politicians prioritize public welfare and policy outcomes over personal or political gain. &  This statement may lead to varied interpretations and might seem subjective, requiring respondents to reflect on personal perceptions of political behavior. The statement also lacks a specific context or scope, making it difficult for participants to make a definitive judgment.\\
        \hline
        Politicians engage in ethical and just behaviors. & Politicians consistently adhere to transparency and accountability in their policy implementations. & The statement is vague and may not effectively discriminate among participant perceptions about what constitutes "ethical and just behaviors," leading to non-specific responses and neutral positions.\\
        \hline
        The government efficiently utilizes public resources. & The government demonstrates optimal allocation and management of public resources for community benefits and economic growth. & This statement is broad and lacks specificity, leading to subjective interpretations of what constitutes "efficient use" and the difficulty in assessing the performance of the government without specific examples.\\
        \hline
    \end{tabular}
    \caption{The statements generated by the Questionnaire generation pipeline (the left column), the same statements after the changes suggested by the LLM-pretesting pipeline (the middle column), and the reasoning from the LLM for that change (the right column) for the Political trust study (RQ1). The other five statements are shown in Table \ref{tab:wvs-llm-pretest}.}
    \label{tab:wvs-llm-pretest-remaining}
\end{table}

\clearpage
\section{RQ2: Cross-cultural study}
\label{appendix:rq2-cross-cultural}
\begin{table}[ht]
    \centering
    \begin{tabular}{|p{5cm}|p{6cm}|p{6cm}|}
       \hline
       \textbf{Question from the standard survey}  & \textbf{Question after LLM-pretesting} & \textbf{LLM reasoning} \\
       \hline
       How much do you support or oppose providing tax rebates for people who purchase energy-efficient vehicles or solar panels? \newline Options: Strongly support, Somewhat support, Somewhat oppose, Strongly oppose & How much would you support financial incentives, such as tax rebates, for people who purchase energy-efficient vehicles or solar panels? \newline Options: Strongly support, Somewhat support, Somewhat oppose, Strongly oppose & The question presupposes that participants understand the concept of tax rebates and its implications. However, it might be too technical for some respondents, particularly if they lack detailed knowledge about fiscal policies. \\
       \hline
       Do you think Congress should be doing more or less to address global warming? \newline Options: Much more, More, Currently doing the right amount, Less, Much less & In the context of your country's government, do you think your national legislative bodies should be doing more or less to address global warming? \newline Options: Much more, More, Currently doing the right amount, Less, Much less & The question is less relevant to participants who may not be familiar with specific governmental structures outside their own country, such as those in South Africa discussing U.S. Congress. \\
       \hline
       How often do you hear about global warming in the media? \newline Options: At least once a week, At least once a month, Several times a year, Once a year or less often, Never & How often do you encounter discussions about global warming in the media, such as news articles, TV reports, or social media? \newline Options: Daily, Several times a week, Once a week, Once a month, Rarely, Never & The answer options are vaguely defined, with interpretations about the frequency of media consumption potentially leading to inconsistencies across respondents’ perceptions of media engagement. \\
       \hline
    \end{tabular}
    \caption{The original questions, the questions after making the changes suggested by the LLM, and the reasoning given by the LLM for the changes. The other three questions are shown in Table \ref{tab:rq2-questions-list}.}
    \label{tab:rq2-questions-list-remaining}
\end{table}

\clearpage
\section{Results}
\begin{table}[ht]
\centering
\caption{Comparison of Traditional/LLM-generated, LLM-Pre-tested mean ratings (with standard deviation), and Mean Differences and p-values for the different metrics based on paired samples t-test. Significant p-values are indicated with an asterisk ($\ast$), and marginally significant p-values (0.05 $\leq$ p < 0.1) are marked with a dagger ($\dagger$).}
\begin{tabular}{lrrrrr}
\toprule
     Metric &  Trad./Gen. score &  Pre-tested Score &  Mean diff. &   Diff. (\%) &  p-value \\
\midrule
     Political Trust study (Prolific, N=238) &   &   &   &   &   \\
\midrule
   \textbf{clarity} &                \textbf{3.16 $\pm$ 1.13} &              \textbf{2.86 $\pm$ 1.10} &                  \textbf{-0.29} &                            \textbf{-9.29} &     \textbf{<0.05*}     \\
     bias &                3.50 $\pm$ 1.06  &              3.48 $\pm$ 1.00 &                -0.02 &                           -0.62 &     0.46 \\
  relevance &                3.62 $\pm$ 1.01 &              3.58 $\pm$ 0.98 &                 -0.03 &                            -0.95 &     0.16     \\
     \textbf{specificity} &                \textbf{2.86 $\pm$ 0.83} &              \textbf{3.00 $\pm$ 0.81} &                \textbf{0.14} &                           \textbf{4.95} &     \textbf{<0.05*}    \\
 
\midrule
     Political Trust study (Experts, N=13) &   &   &   &   &   \\
\midrule
   \textbf{clarity} &                \textbf{3.61 $\pm$ 1.16} & \textbf{2.95 $\pm$ 1.27} &                  \textbf{-0.66} &                            \textbf{-18.34} &     \textbf{<0.05*}     \\
     \textbf{bias} &                \textbf{2.83 $\pm$ 1.16} & \textbf{3.13  $\pm$ 1.24} &                \textbf{0.30} &                           \textbf{10.60} &     \textbf{<0.05*} \\
  relevance &                4.17 $\pm$ 0.88 &              3.94  $\pm$ 1.01 &                 -0.23 &                            -5.54 &     0.05-0.1†     \\
     \textbf{specificity} &                \textbf{2.98  $\pm$ 1.07} &              \textbf{2.72  $\pm$ 0.94} &                \textbf{-0.26} &                           \textbf{-8.79} &     \textbf{<0.05*}   \\
 
\midrule
     Cross-cultural study (Prolific, N=118) &   &   &   &   &   \\
\midrule
   clarity &                3.87 $\pm$ 1.12 &              3.97 $\pm$ 1.07 &                  0.1 &                            2.47 &     0.05-0.1†    \\
     bias &                2.58 $\pm$ 1.01 &              2.51 $\pm$ 1.03 &                -0.08 &                           -2.94 &     0.05-0.1† \\
  relevance &                3.88 $\pm$ 1.13 &              3.93 $\pm$ 1.11 &                 0.05 &                            1.33 &     0.26     \\
     specificity &                3.77 $\pm$ 0.87 &              3.72 $\pm$ 0.93 &                -0.05 &                           -1.35 &     0.15    \\
 
\bottomrule
\end{tabular}
\label{tab:summary}
\end{table}

\begin{table}[h]
    \centering
    \resizebox{0.95\textwidth}{!}{
    \begin{tabular}{p{4cm} p{4.5cm} p{4.5cm} p{4.5cm}}
        \toprule
        \textbf{Theme} & \textbf{RQ1 (Participants) Example Quote} & \textbf{RQ1 (Experts) Example Quote} & \textbf{RQ2 Example Quote} \\
        \midrule
        Clarity and Readability & LLM-generated (G1) is preferable because it is more concise. Although LLM pre-tested (G2) is a better question, it is not as clear. & Shorter sentences -> more understandable. & The first one is clearer because it implies that conversations don't have to be in-depth to count. \\
        \midrule
        Specificity and Detail & LLM-generated (G1) questions are far less specific than LLM pre-tested (G2) questions. & Although the statements in LLM pre-tested (G2) seem more precise, they are more difficult to answer for the general public because they are too specific. & Q2 is more detailed and clearer by giving the extent of the discussions. \\
        \midrule
        Bias or Fairness Concerns & Both statements in LLM pre-tested (G2) are too convoluted in terms of the adjectives used to describe various nouns, making the statements a little leading. & LLM pre-tested (G2) questions ask about two things simultaneously, so aren't clear. & Because question 2 is biased by the word 'in-depth'. \\
        \midrule
        Relevance and Applicability & It mentions the government’s record. LLM-generated (G1) just says actions, while LLM pre-tested (G2) gives specifics. & The reading age for LLM pre-tested (G2) questions is too high for a general audience. & Not all discussions may get to the point of it being in-depth but the point of having a discussion is still relevant. \\
        \midrule
        Conciseness and Simplicity & The statements are more straightforward and easier to understand than LLM pre-tested (G2). & LLM-generated (G1) is clearer because it uses straightforward and concise language that is easier to understand. & They are basically saying the same thing, however, question one is shorter and therefore needs less time to understand. \\
        \bottomrule
    \end{tabular}}
    \caption{Thematic Analysis Summary for RQ1 and RQ2}
    \label{tab:thematic_analysis}
\end{table}

\end{document}